\documentclass{elsart3}

 \usepackage{graphicx}

\usepackage{amssymb}
\usepackage{subfigure}

\begin{document}

\begin{frontmatter}



\title{Qualification Procedures of the CMS Pixel Barrel Modules}

\author[eth]{A. Starodumov\corauthref{cor1}}\ead{Andrey.Starodumov@cern.ch},
\author[psi]{W. Erdmann}, 
\author[psi]{R. Horisberger}, 
\author[psi]{H.Chr. K\"{a}stli}, 
\author[psi]{D. Kotlinski}, 
\author[eth]{U. Langenegger}, 
\author[eth]{B. Meier},
\author[psi]{T. Rohe}, 
\author[eth,psi]{P. Tr\"{u}b}
\address[eth]{Institut f\"{u}r Teilchenphysik, ETH-Z\"{u}rich, 8093 Z\"{u}rich, Switzerland}
\address[psi]{Paul Scherrer Institut, 5232 Villigen PSI, Switzerland}
\corauth[cor1]{Corresponding author}

\begin{abstract}
 The CMS pixel barrel system will consist of three layers built of about 800 modules. One module contains 66560 readout channels and the full pixel barrel system about 48 million channels. It is mandatory to test each channel for functionality, noise level, trimming mechanism, and bump bonding quality. Different methods to determine the bump bonding yield with electrical  measurements have been developed. Measurements of several operational parameters are also included in the qualification procedure. Among them are pixel noise, gains and pedestals. Test and qualification procedures of the pixel barrel modules are described and some results are presented.
\end{abstract}

\end{frontmatter}

\section{Introduction}
\label{1}
The CMS detector will be built at the Large Hadron Collider (LHC) at CERN. A hybrid pixel system will be the detector closest to the interaction point and should provide reliable pattern recognition, efficient and precise momentum and vertex reconstruction in an extremely high track density environment. This dictates a fine granularity that correspondingly leads to a large number of readout channels. The testing of the detector building blocks---modules---of such a system becomes a challenge. A complete description of the CMS pixel detector and its design principles are given elsewhere \cite{CMS}. Here, some details relevant to the qualification of pixel modules will be mentioned. 

The pixel barrel system will be composed of three cylinders with radii of 4, 7 and 11\,cm and a length of 52\,cm. Each cylinder is built of ladders with 8 modules placed on it. To assemble the pixel barrel detector about 800 modules will be needed. The pixel barrel module consists of a single sensor substrate with 16 front-end readout chips (ROC) bump-bonded to it and a hybrid circuit (HDI---high density interconnect) mounted on top of the sensor. Two thin strips of Si$_{3}$N$_{4}$ glued to the readout chips serve as a base to attach the module to the cooling frame.
  
The whole pixel barrel detector will contain about 48 million readout channels. It is mandatory to test each channel for functionality, noise level, trimming mechanism, and bump bonding quality. The qualification process also includes the determination of the operational parameters (like trim bit settings, measurement of noise, gains and pedestals), a check of the sensor $I$-$V$ dependence  and a thermal cycling test. The time scale for the barrel detector construction is about one year. This implies a necessity to test four modules a day. To fulfill this time-constraint it is anticipated to use only tested components and perform a failure diagnostics in parallel with the qualification tests. Further details about the assembling procedure of the pixel modules can be found in \cite{stefan}.

In the following qualification tests and procedures will be described. Some test results will be presented to illustrate the quality of  ROCs and modules and to describe qualification criteria established so far. 

\section{Test procedures}
\label{3}
Module qualification implies a thorough check of its performance. Each ROC should be programmable via setting corresponding DAC registers. Every pixel readout circuit has to exhibit a proper behavior. And finally, one has to be able to set each module in the operational regime and calibrate it. Therefore, the test procedure is divided into three main steps.
 
First, all 26 DACs are set to their default values and then the most crucial ones are tuned individually for each ROC. Among them are DACs which control the amplitude of an internal calibrate signal (the corresponding DAC is called Vcal), the analog current (Vana), the comparator threshold (Vthrc) and the delay of the internal calibrate signal with respect to the trigger signal (CalDel). 
 
In the second step the functionality of the pixel readout circuits and their electrical connections to the sensor pixel are checked. The following procedures are foreseen:
\begin{itemize}
 \item {{\bf Pixel test}: check the proper response of each pixel and find defective ones.}
 \item {{\bf Trim bit test}: control the functionality of the four trim bits responsible for fine tuning of each pixel's threshold.}
 \item {{\bf Bump-bonding test}: control the bump-bonding quality and find missing or defective bumps.}
 \item {{\bf Pixel address test}: verify that each pixel readout circuit responds with the correct pixel address.} 
\end{itemize}

Finally, it is necessary to determine the main characteristics, which allows to set a module to the operational regime and calibrate it, and to validate proper functionality of the module under operational voltages, working temperature and temperature variations. This is done by performing the following procedures:
\begin{itemize}
 \item{{\bf Noise}: determine pixel noise by measuring the S-curve for each pixel.}
 \item{{\bf Trimming}: setting thresholds of each pixel to obtain a uniform response over a whole module.}
 \item{{\bf Gains and pedestals}: perform a final calibration of each pixel.}
 \item{{\bf $I$-$V$ curve}: verify the absence of the sensor breakdown and the high leakage current.}
 \item{{\bf Thermal cycle}: test the module behavior under different temperatures and determine all relevant parameters for the working temperature (-$10^{\circ}$C).}
\end{itemize}

Some tests and procedures described in the following are specific for the CMS pixel detector readout chip. Further details about the ROC design can be found in \cite{hans}.

\subsection{Pixel test}
\label{4}
For each pixel the functionality is checked by inducing a signal via an internal calibrate capacitance. First, it is controlled that the masked (disabled) pixel does not respond if the calibrate signal is sent to it. Second, for the enabled pixel it is verified that if a single calibrate signal is sent one and only one output signal is registered. As a result of this test a list of defective pixels is produced. Three modules have been tested so far and only 6 dead pixels have been found out of almost 200000 pixels.

\subsection{Trim bit test}
\label{5} 
To fine tune the threshold for each pixel four trim bits are used. The functionality of each trim bit for every pixel is controlled during this test. Thresholds are measured for each pixel five times: 1) for untrimmed state, 2)-5) for trimmed states, when each time only one trim bit is enabled. Trimmed and untrimmed thresholds are then compared per pixel. If a certain bit does not work, the measured thresholds for trimmed and untrimmed state are equal. The addresses of pixels with defective trim bits and the faulty trim bit number are recorded. Nary defective trim bit has been found in the three tested modules.
   
\subsection{Bump-bonding test}
\label{6}

\begin{figure}[t]
\includegraphics[scale=.35]{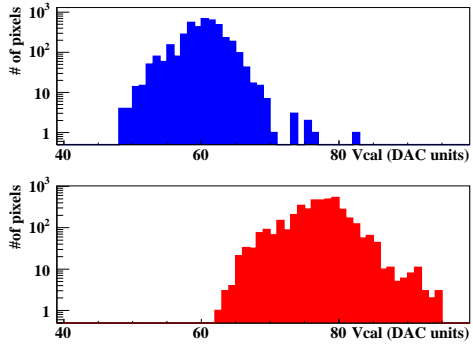}
\caption{Pixel threshold distribution for a single ROC taken with an external calibrate signal (top) and via cross-talk mechanism (bottom).
\label{fig:bb}}
\end{figure}

\begin{figure}[b]
\includegraphics[scale=.35]{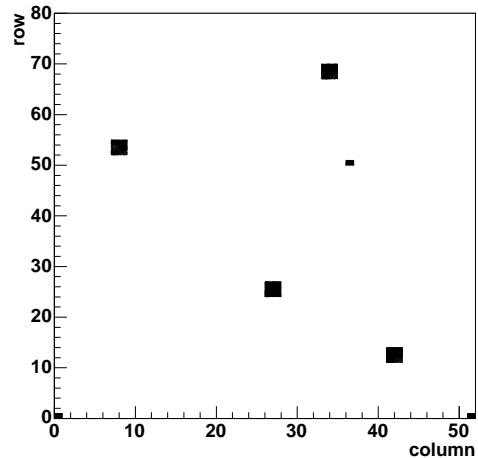}
\caption{Threshold difference map ($52 \times 80$ pixels) taken with the `modified external calibration' method applied to a ROC with removed (colored in black) bump bonds.}
\label{fig:bbmap}
\end{figure}

A bump bonding procedure has been developed at PSI (for details see \cite{rohe}). The first bump bonding test will be done on bare modules. But since bonds can be damaged during the module assembly it is mandatory to repeat the bump bonding test to identify pixels with missing or broken bumps in the fully equipped modules. To speed up and simplify the procedure several electrical methods without radioactive sources have been developed. Two of them rely on the fact that if the ROC preamplifier is set close to saturation and a high leakage current is drawn through the bump, the preamplifier saturates. If the bump is missing, the preamplifier is not saturated. A high leakage current is generated with a light source (a lamp, for example) or with a positive bias. Any of these two methods can be used with bare modules when the sensor is not yet covered by the HDI. 

For the completely assembled module another method is applied. This method is called the `modified external calibration'. In the ROC a possibility to send a calibrate signal through the sensor is implemented (see \cite{hans}). In principle, this functionality allows the identification of missing bumps by measuring pixel thresholds. However, due to cross-talk in the chip even for pixels with missing bumps the threshold values are close to values measured for pixels with bumps (see fig. \ref{fig:bb}). However, the cross-talk effect can help to identify missing bumps. Taking two threshold maps---with external calibrate signal and via cross-talk---and then comparing them, one can find that pixels with missing bumps have the same threshold in both measurements. In such a way missing or defective bumps are identified and their position is recorded for the final module qualification.

The procedures have been validated by applying them to several specially prepared ROCs with a sensor where a few bumps have been removed manually before bump bonding (see fig. \ref{fig:bbmap}). Measurements have shown 100\% efficiency for missing bump finding. 
In one tested module no broken bumps have been found. In two other modules 4 and 7 broken bumps have been detected. Most of them are located at the edges in the first or last columns.

\subsection{Pixel address test}
\label{7}

\begin{figure}[b]
\includegraphics[scale=.35]{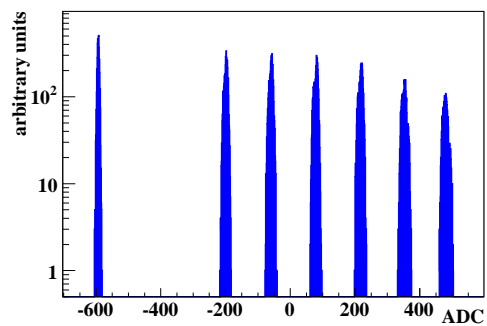}
\caption{`Ultra black' (left most) signal and six levels of the analog signal coded pixels addresses.\label{fig:add}}
\end{figure}
An individual pixel address consists of five analog signals: two signals are devoted to the column index and three to the row index. Each address is coded into a six level analog signal. To decode correctly the pixel address, these levels have to be well separated. The pixel address decoding test has two goals: first, to measure the six levels and find the separation values, and second, to decode the address of each pixel and compare with the true one. Fig. \ref{fig:add} shows six levels and the `Ultra black' signal level (which is used to mark the beginning and the end of the data packet and to separate individual ROC information). All levels in the figure are well separated from each other and for such a ROC there is no problem to decode pixel addresses correctly. It is also possible that some individual pixels generate a wrong address, then such pixels are counted as defective. In the three tested modules no wrong addresses have been observed.   

\subsection{Noise}
\label{8}
\begin{figure}[t] 
\includegraphics[scale=.35]{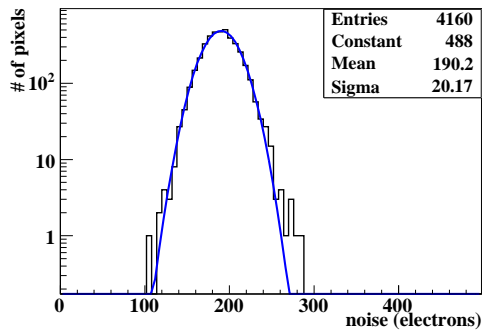}
\caption{A noise (in electrons) distribution for a single ROC.\label{fig:noise}}
\end{figure}

The threshold for each pixel in a module should be determined and noisy pixels should be disabled. The threshold is supposed to be set at least 5 times higher than the average pixel noise in a ROC. Pixels with noise 4-5 times higher than the average will flood the readout system with a high rate of hits and cause significant dead time and data losses. Therefore such pixels should be masked. A method to determine the noise level of a pixel is to take an S-curve, which is the efficiency versus the amplitude of the calibrate signal. The width of the S-curve provides the noise measurement, the fit is done using the {\it error function}. The result of such a test for a single ROC is presented in fig. \ref{fig:noise}. The average noise and the spread of the noise distribution also influences the module qualification. For non irradiated detectors the threshold will be set at 2000-2500 electrons, hence the average noise value should not exceed 500 electrons and the spread should not be more than a few hundred electrons.

\subsection{Trimming}
\label{9}

\begin{figure*}[t]
\subfigure[]{\includegraphics[scale=.35]{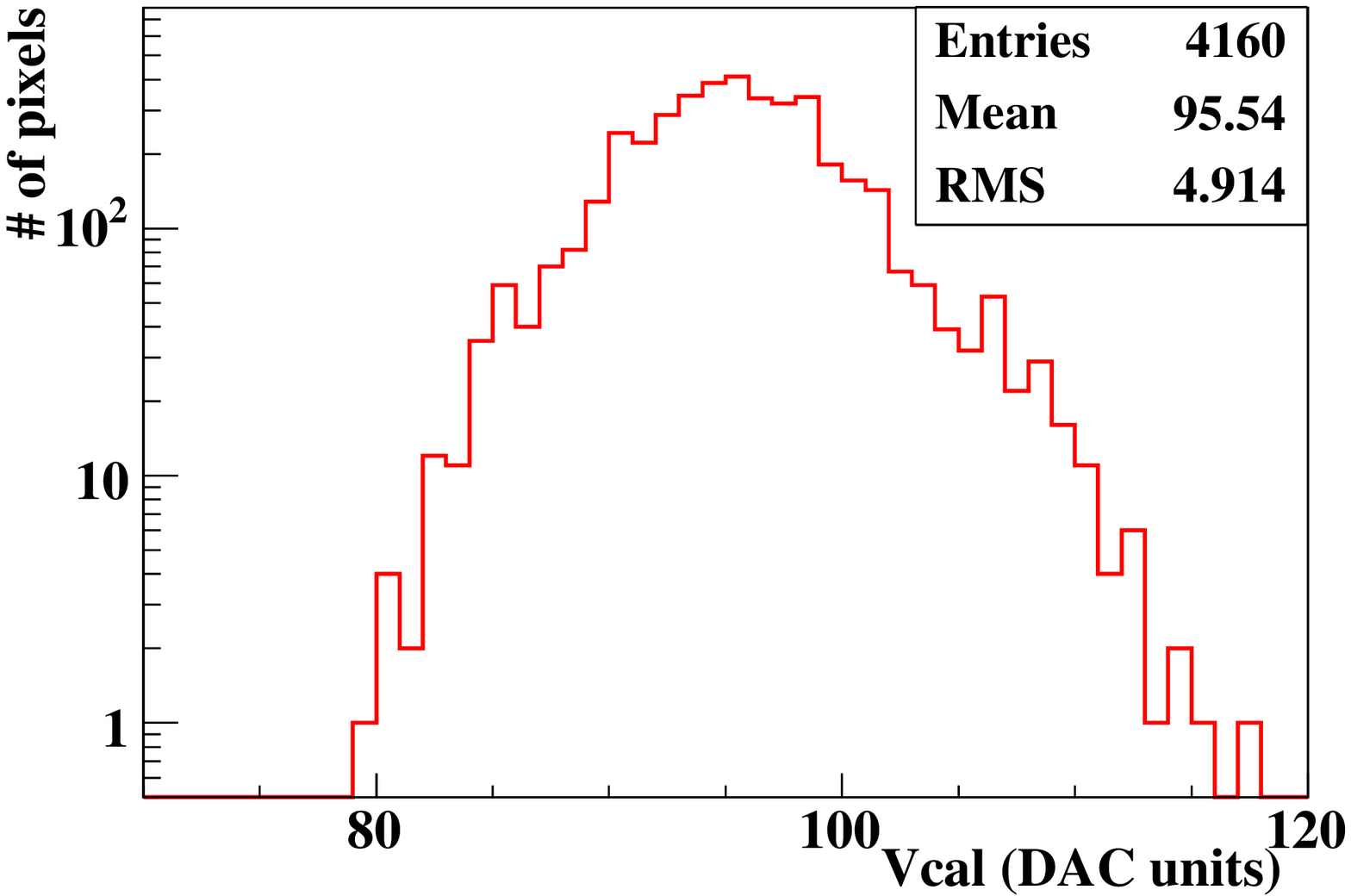}}
\hfill
\subfigure[]{\includegraphics[scale=.35]{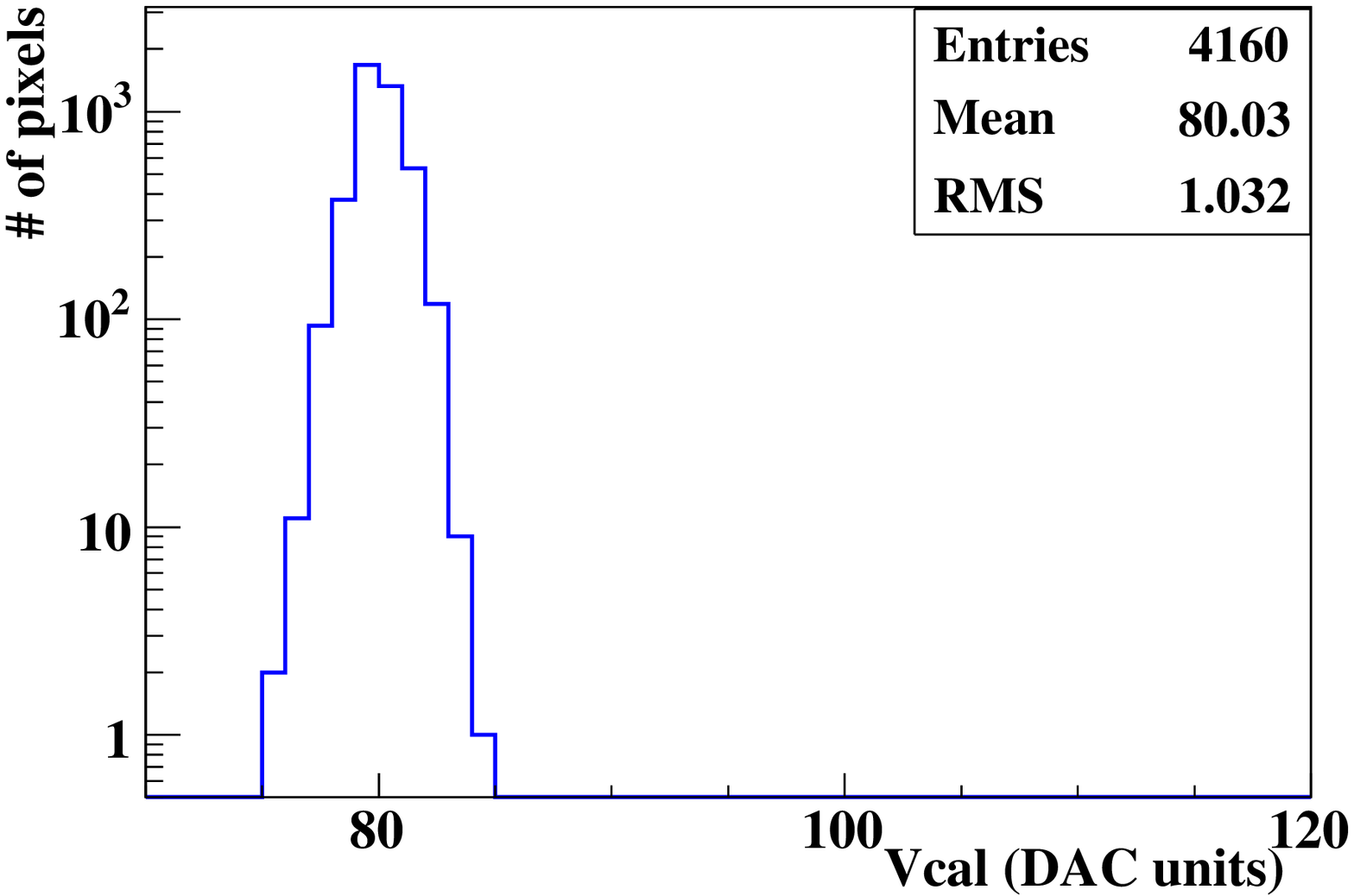}}
\caption{Pixel threshold distribution for untrimmed (a) and trimmed (b) readout chip.\label{fig:trim}}
\end{figure*}

The aim of the trim algorithm is to unify the response of all pixels in a ROC. In practice this means the unification of the pixel's thresholds. A common global threshold for all pixels can be set per ROC (Vthrc DAC). To account for the pixel to pixel variations a trim mechanism is implemented with four trim bits which can be set in each pixel. This mechanism allows to set 16 trim states. By setting trim bits the threshold of the pixel is decreased. The size of the correction is determined by the trim voltage, which is set per ROC via a DAC called Vtrim. 

The only relevant input parameter to the algorithm described below is the effective threshold at which the response is to be unified. This threshold is fixed by choosing the value of the Vcal DAC, which controls the height of the internal calibrate pulse. The Vcal value can be related to the charge expressed in electrons. 

The first step is to find a value for the global threshold which corresponds to the target Vcal value. This is done by measuring for each pixel the global threshold value at which it starts to respond (with 50\% probability). Since the effective thresholds can only be lowered afterwards, the maximum value of this distribution determines the common global threshold per ROC. The found Vthrc value is used during the rest of the algorithm and recorded as a parameter for the trimmed ROC.

The second step of the trim algorithm is to find an appropriate trim voltage. First, the Vcal value is measured for each pixel at the threshold fixed on the previous step. The pixel, which needs the highest Vcal value is used to determine the Vtrim. For this pixel all trim bits are turned on. Then the trim voltage is increased, until the Vcal of this pixel reaches the target value. This Vtrim value is the second parameter set per ROC. 

In the third step of the algorithm trim states are set for each pixel. It is done by finding the trim bit configuration at which pixels start to respond to the target Vcal value. In fig. \ref{fig:trim} Vcal distributions are shown for an untrimmed and a trimmed ROC. The target Vcal value in this example is set to 80. The uniformity is improved by a factor of 5 after trimming. The achieved uniformity (the r.m.s. value in fig. \ref{fig:trim}(b)) is about 70 electrons. 

\subsection{Gains and pedestals}
\label{10}
\begin{figure*}
\subfigure[]{\includegraphics[scale=.35]{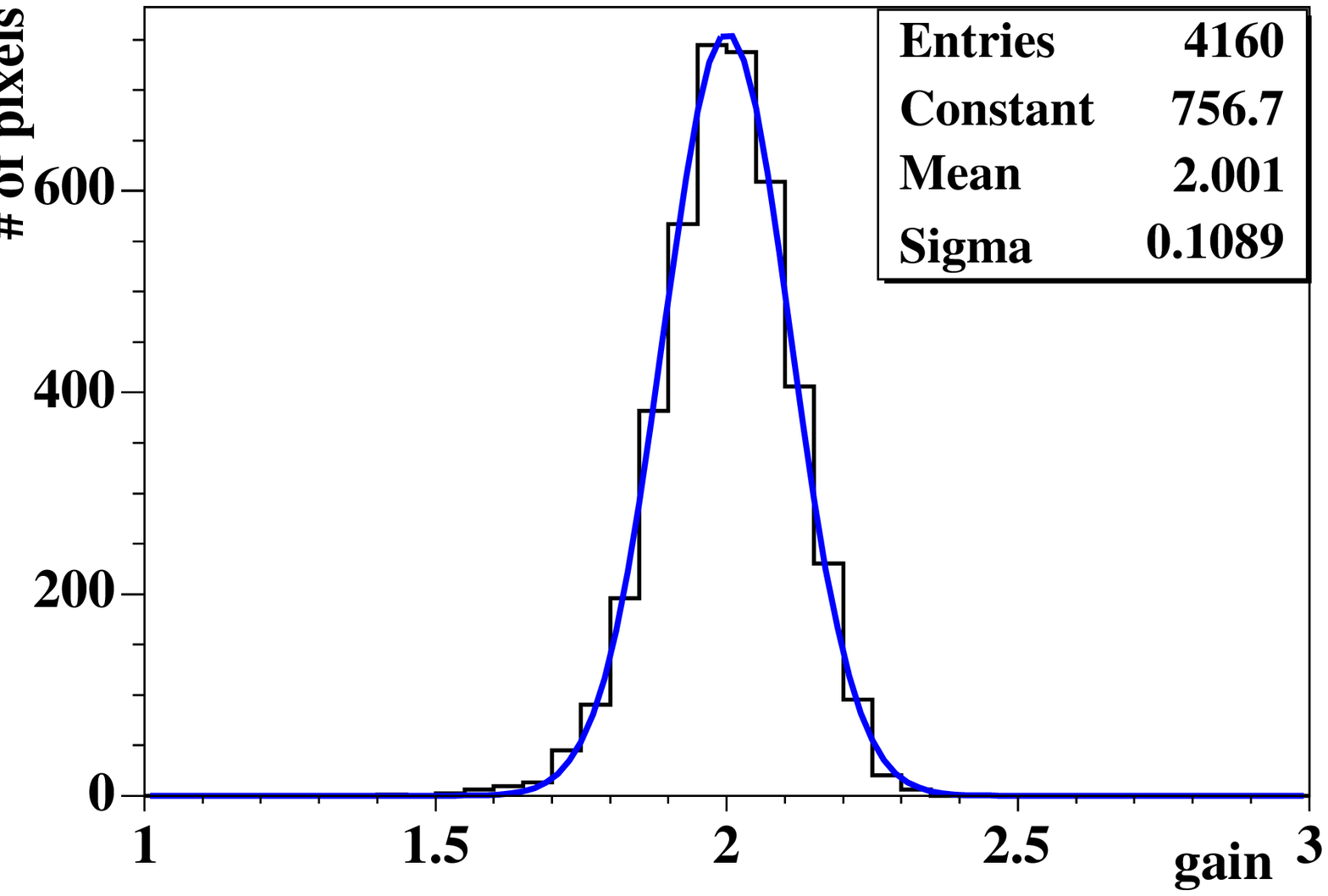}}
\hfill
\subfigure[]{\includegraphics[scale=.35]{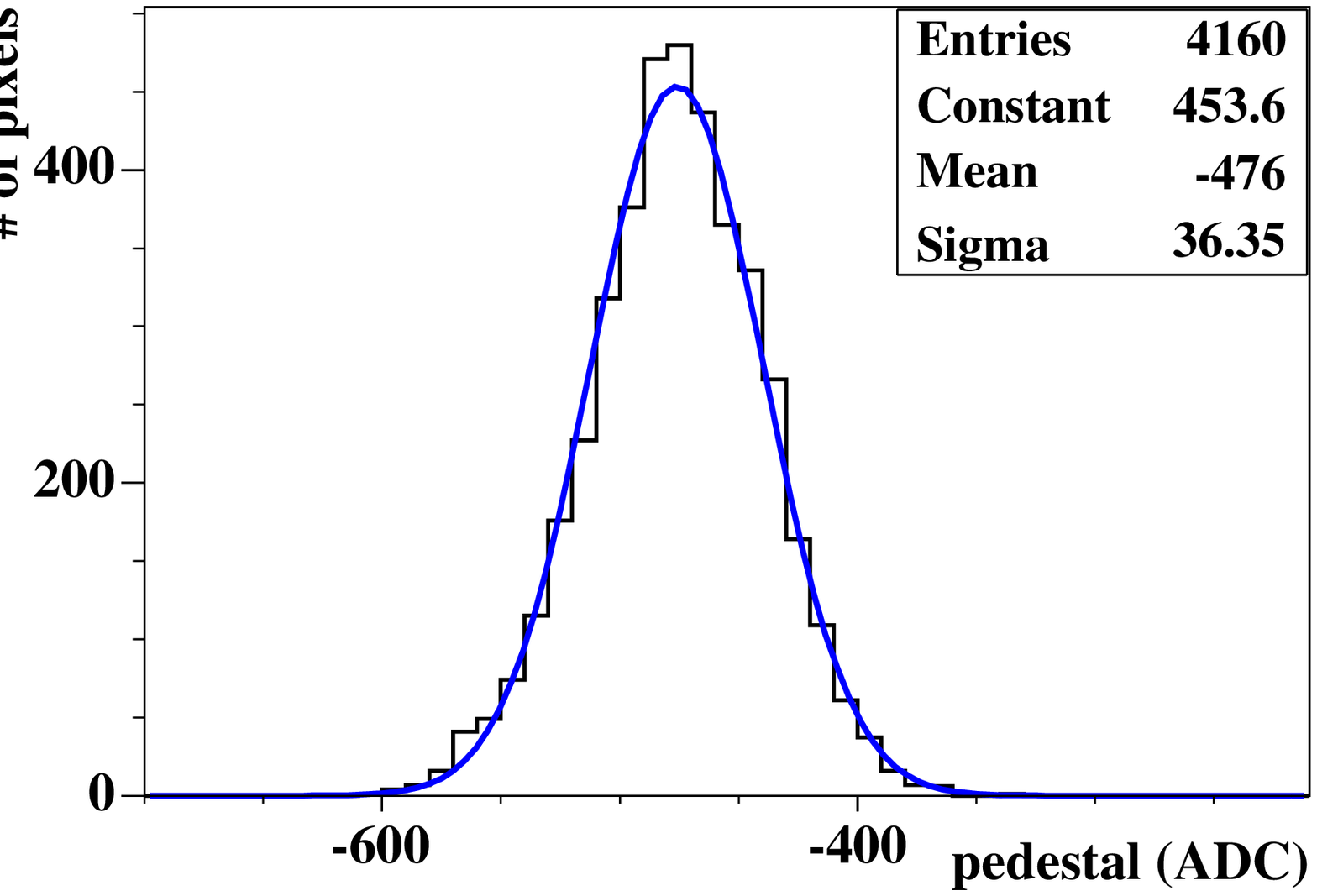}}
\caption{Gain (a) and pedestal (b) distributions for a single ROC\label{fig:gain}}
\end{figure*}
Another qualification item is the determination of gains and pedestals for each pixel. The gains and pedestals are used to convert the charge collected by pixels and measured by ADC counts to electrons. The calibration parameters are needed in off-line and on-line (at high level trigger) analysis. It may be problematic to have a precise on-line calibration  based on individual pixels because of their large number. An obvious solution would be the use of approximate calibration constants averaged over one ROC. This will result in a degradation of the hit resolution because of not precise charge interpolation. The use of approximate values of gains and pedestals for all pixels in a ROC implies that the variations of these parameters among pixels cannot be too large. The spread in the parameters is acceptable if the mis-calibration contribution to the track and vertex reconstruction precision is less then multiple scattering and misalignment effects. According to \cite{danek} the tolerable variation of gains is about 20\%-40\% and the pedestal variation might be as large as 1000-2000 electrons.     

The test is performed by injecting various amplitudes of the calibrate signal and measuring the ADC response. The resulting distribution is fitted by a linear function and the slope (gain) and offset (pedestal) are recorded for each pixel. These distributions provide twofold information. If any of the distributions is too broad the module should be rejected. Also, if some pixels have a gain and/or a pedestal far (e.g. more than 4-5 times the r.m.s.) from the mean value, these pixels are counted as defective ones. In fig. \ref{fig:gain} the distributions of gains and pedestals are presented for a single ROC. The standard deviation of the gain distribution is about 5\% and of the pedestal distribution is about 1200 electrons.        

\subsection{I-V curve}
\label{11}
The measurement of the sensor leakage current versus the depletion voltage ($I$-$V$ curve) will be done several times during the module production. Finally this measurement will be performed on a fully assembled module with the aim to control the absence of any global individual pixel damage which might occur during assembly. The voltage will be varied from 0 to 600\,V. The total leakage current should not exceed 2\,$\mu$A at the operational voltage ($V_{OP}$) of 150\,V. The variation of the leakage current should satisfy the following constraint $I(V_{OP}) / I(V_{OP} - 50V) \le 2$.     

\subsection{Thermal cycling}
\label{12}
\begin{figure}
\includegraphics[scale=.49]{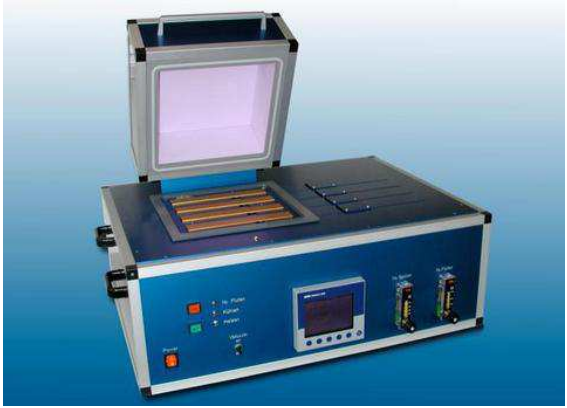}
\caption{A set-up for thermal test of CMS pixel barrel modules\label{fig:temp}}
\end{figure}

The CMS pixel detector will be operated at a low temperature of $-10^{\circ}$C. To verify the proper performance at this temperature a thermal test is foreseen during the qualification procedure. In fig. \ref{fig:temp} a set-up for a thermal cycling is shown. Up to four modules can be tested concurrently. The thermal test will last about 24 hours and the temperature will be varied from $+30^{\circ}$C to $-10^{\circ}$C about 10 times. Operational parameters (like trim bits, gains and pedestals) will be obtained and recorded. Any failure during this procedure will eliminate the use of the module in the final construction.

\section{Module qualification}
\label{13}
Based on the results of all tests described above the modules will be qualified for use in the CMS pixel barrel detector. A grading scheme is currently under development. Only after having tested a reasonable amount of modules one can completely define the qualification criteria. For the time being many questions remain open, like the definition of the defective pixels. This is clear for pixels not responding or pixels with missing bump. It is not so obvious in the case of defective trim bits (one or two), highly resistive bump bonds (which still work) or broad pedestal distribution. Generally, modules will be sorted in three or four classes of quality. Those which pass the quality tests and have less than 1\% of defective pixels will be qualified to be used in the pixel system. If the amount of defects is between 1\% and 2\%,  modules may be considered as spare ones. If the number of defective pixels is more than 2\%, modules will be rejected. In the three modules tested so far the maximum fraction of defective pixels is less than $10^{-4}$. 

\section{Conclusion}
\label{14}
In the coming years about 800 pixel modules will be assembled at PSI. Each of them should pass comprehensive tests and be qualified to be used in the construction of the CMS pixel barrel detector. A qualification procedure has been established to ensure a reliable and high-quality device. One of the most crucial tests is the bump bonding quality. Several procedures have been developed and validated. All of them provide consistent results. Another important procedure is the trimming of the ROCs. A sophisticated but fast algorithm has been developed to guarantee an excellent unification of the pixel thresholds down to 2\%. The measurement of the pixel noise, gain and pedestal allows to set a module in the correct operational regime. $I$-$V$ test and thermal cycling  procedure ensure that modules can be operated under CMS conditions. The overall qualification procedure will be tuned and verified during the module pre-production period. 

\section{Acknowledgment}
\label{15}
The authors would like to express gratitude to all colleagues from the Laboratory of Particle Physics at PSI who helped us to build a test setup, shared their knowledge of the pixel detector, discussed test procedures and results.
 
We thank K. Gabathuler of PSI for his attention to our everyday needs, for providing us with necessary documentation about CMS pixel barrel detector in general and ROC in particular.  

We appreciate fruitful discussions with Ch. H\"{o}rmann of the University of Z\"{u}rich and PSI and S. K\"{o}nig of PSI.

Finally, our special thanks to S. Streuli of ETH Z\"{u}rich who has not only built all the necessary equipment, especially the devices with missing bumps but from whom we learned many useful things.



\end{document}